\documentclass[12pt,a4paper]{article}
\usepackage{a4wide}
\usepackage{latexsym}
\usepackage{epsf}
\usepackage{amssymb}
\begin{document}
\def\be{\begin{equation}}
\def\ee{\end{equation}}
\def\bea{\begin{eqnarray}}
\def\eea{\end{eqnarray}}

\def\pd{\partial}
\def\a{\alpha}
\def\b{\beta}
\def\g{\gamma}
\def\d{\delta}
\def\m{\mu}
\def\n{\nu}
\def\t{\tau} 
\def\l{\lambda}
\def\s{\sigma}
\def\e{\epsilon}
\def\scri{\mathcal{J}}
\def\cM{\mathcal{M}}
\def\tcM{\tilde{\mathcal{M}}}
\def\RR{\mathbb{R}}
\def\CC{\mathbb{C}}

\hyphenation{re-pa-ra-me-tri-za-tion}
\hyphenation{trans-for-ma-tions}


\begin{flushright}
IFT-UAM/CSIC-01-02\\
hep-th/0101181\\
\end{flushright}

\vspace{1cm}

\begin{center}

{\bf\Large Non-critical Non-singular Bosonic Strings, Linear Dilatons 
and Holography}

\vspace{.5cm}

{\bf Enrique \'Alvarez},
{\bf  C\'esar G\'omez ${}^{\spadesuit}$},
{\bf Lorenzo Hern\'andez},
{\bf and Pedro Resco} \\
\vspace{.3cm}

\vskip 0.4cm

{\it Instituto de F\'{\i}sica Te\'orica, C-XVI,
  Universidad Aut\'onoma de Madrid \\
  E-28049-Madrid, Spain}\footnote{Unidad de Investigaci\'on Asociada
  al Centro de F\'{\i}sica Miguel Catal\'an (C.S.I.C.)}

and

{\it Departamento de F\'{\i}sica Te\'orica, C-XI,
  Universidad Aut\'onoma de Madrid \\
  E-28049-Madrid, Spain }

\vskip 0.2cm

${}^{\spadesuit}$\ {\it I.M.A.F.F., C.S.I.C., Calle de Serrano 113\\ 
E-28006-Madrid, Spain}

\vskip 1cm


{\bf Abstract}
\end{center}
$AdS_{5}$ with linear dilaton and non vanishing  $B$-field is
shown to be a solution of the non critical string beta function equations.
A non critical $(D=5)$ solution interpolating between flat space-time and
$AdS_{5}$, with asymptotic linear dilaton and
non vanishing  $B$-field is also presented. This solution is free of 
space-time singularities and
has got the string coupling constant everywhere bounded. 
Both solutions admit holographic interpretation in
terms of ${\cal N}=0$ field theories. Closed string tachyon stability is 
also discussed.


\begin{quote}

\end{quote}


\newpage

\setcounter{page}{1}
\setcounter{footnote}{1}

\section{Introduction}
Much effort has recently been devoted to research in Anti
de Sitter (AdS) string backgrounds. The reason is at least,
twofold.
First of all, holography (cf. \cite{maldacena}, \cite{gubser}\cite{witten})
relates bulk gravitational physics in $AdS_5$ with some
conformal field theory (CFT) at the boundary.
In addition, AdS is by far the simplest geometry where the
ideas of Randall-Sundrum \cite{randallsundrum}  on 
four-dimensional confinement of gravity
and hierarchy generation can be implemented.
\par
The ten-dimensional background $AdS_5\times S_5$ is a well-known
background geometry for the type $IIB$ string description of the near horizon
limit of D3 branes. In order for it to become a consistent string background,
a self-dual Ramond-Ramond (RR) five-form has to be turned on. The five sphere
$S_5$ plays the r\^ole of an internal manifold, and is associated with the 
extra scalars required by ${\cal{N}}=4$ supersymmetry.
\par
One of the main motivations of the present paper stems from the natural
question as to whether $AdS_5$ can define a non-critical string
background. We will actually show this to be the case with euclidean signature,
provided both an appropiate
dilaton and Kalb-Ramond field are turned on.
\par
A simple extrapolation of the holographic principle to the present situation
suggests that this background should be dual with some four-dimensional,
non-supersymmetric,
${\cal{N}}=0$, CFT on the boundary.
\par
Actually, the simplest non-critical bosonic string background is the 
{\em linear dilaton} \cite{myers}, with a dilaton field:
\be\label{ld}
\Phi = \pm Q r
\ee
living in flat $d$-dimensional Minkowski (or euclidean) spacetime:
\be
ds^2=d\vec{x}_{(1,d-1)}^2 + dr^2
\ee
with
\be
q\equiv\frac{d-25}{6\a'}\equiv - Q^2
\ee
(where $Q\in \mathbb{R}$ whenever $D\equiv d+1 <26$).
\\
It is plain that we actually have two $\mathbb{Z}_2$-related different potential vacua
corresponding to the two admissible signs in (\ref{ld}). Bearing in mind a holographic
interpretation of the coordinate $r$, it is natural to ask
whether there are topologically non-trivial configurations which interpolate between
these two vacua, i.e., with a dilaton behavior of the type
\bea
&&\Phi(r)_{r\rightarrow \infty}\sim -  Q r\nonumber\\
&&\Phi(r)_{r\rightarrow -\infty}\sim +  Q r
\eea
The answer is in the affirmative (again, only with euclidean signature). The first
interpretation of this solution that comes to mind is that of a sort
of  tunneling between the two $\mathbb{Z}_2$-related linear dilaton behaviors
defined above.
\par
Asymptotically $({r\rightarrow \infty})$ this interpolating solution is just
the linear dilaton in flat space-time with weak string coupling constant. At 
${r\rightarrow -\infty}$ the solution becomes $AdS_{5}$ with linear dilaton,
 but
again with weak string coupling constant ( see Fig 2 ). In addition the 
solution is
free of singularities with a bounded curvature everywhere.
\par
It is of course interesting to look for the holographic interpretation of
this solution. 
\par
String backgrounds that behave asymptotically at 
${r\rightarrow \infty}$
as
\bea\label{lid}
&&ds^2=d\vec{x}_{(1,d-1)}^2 + dr^2 + ds^{2}(M)\\
&&g_{s}^{2}=e^{-\Delta r}
\eea
where $\Delta$ is a constant and $M$ a compact internal manifold, 
trivially fibered
on the flat $d+1$ dimensional space-time, have been first considered, 
from the holographic
point of view, in reference \cite{Aharony:1998ub}. 
\par
It was there argued that this type
of string backgrounds are holographic. This fact can be easily understood 
by considering
the metric in the Einstein frame. In this frame ,and thanks to
the linear dilaton, the $d+1$ dimensional space-time metric becomes 
effectively $AdS$ metric, 
which in particular implies the necessary condition of holography, 
namely that timelike
geodesics never reach the boundary. According to \cite{Aharony:1998ub} the holographic 
dual 
of these backgrounds is a non local little string theory.
One of the main problems of linear dilaton backgrounds of type (\ref{lid})
considered in \cite{Aharony:1998ub} is 
how to regulate the divergences in the strong coupling region 
$({r\rightarrow -\infty})$.
\par
At this point it is worth noticing that the interpolating solution we
present in this paper behaves asymptotically exactly as above
(\ref{lid}), up to the fact that we have not any internal compact
manifold\footnote{ Although we could, just by introducing toroidal
  {\em spectator dimensions}}. Moreover our solution automatically
regulates the divergence at ${r\rightarrow -\infty}$ by becoming, in a
smooth way, $AdS_{5}$ with weak string coupling constant. From the
holographic point of view the boundary theory could be a non local little string theory with ${\cal N}=0$ supersymmetry.  If this is
the case it would be very interesting to understand if this
hypothetical non locality alluded to above could be due in our case to
the non vanishing expectation value for the Kalb-Ramond field.
\par
The other solution we find, namely $AdS_{5}$ with linear dilaton and
B-field is also interesting by itself, from the point of view of
holography. The most natural candidate for the boundary theory would
be of course a ${\cal N}=0$ theory coupled to non vanishing B-field. In the context of ${\cal N}=8$ gauged supergravity there are two non trivial $AdS_{5}$ minima associated with expectation values for the $\underline{10}$ and the $\underline{20}$ multiplets \cite{girardello}.
Holographically these two vacua should correspond to ${\cal N}=0$ four
dimensional conformal field theories. In these cases the string
background is critical $AdS_{5}\times W_{5}$ but with the internal manifold
$W$ fibered in a non trivial way on the holographic coordinate. It
would be interesting to study if there exist any relation between
these ${\cal N}=0$ backgrounds predicted by ${\cal N}=8$ supergravity and the
$AdS_{5}$ solution with non vanishing B described in this paper.
\par
Finally let us just mention that the interpolating solution we present is only
valid with euclidean signature. This is the reason we tend to interpret it
as a sort of tunneling effect.

\section{$AdS_5$ and the Bosonic String}
The first step in determining whether strings can live in a given background 
is to check the vanishing of the Weyl anomaly coefficients ,
 that is:

\bea\label{weyl}
&&R_{AB}+ 2 \nabla_A\nabla_B \Phi -\frac{1}{4} H_{ACD}H_{B}^{CD}=0\nonumber\\
&&-\frac{1}{2}\nabla^C H_{CAB}+\nabla^C \Phi H_{CAB}=0  \nonumber\\
&&q+(\nabla\Phi)^2 -\frac{1}{2}\nabla^2 \Phi -\frac{1}{24}H^2 =0
\eea
We have denoted by capital latin letters spacetime indices 
($A,B,\ldots =0,1\ldots D-1\equiv d$), and by greek letters Poincar\'e indices
($\mu,\nu,\ldots = 0,\ldots,d-1$).
\\
We intend to look for backgrounds in which the metric part enjoys Poincar\'e
invariance, that is
\be
ds^2= a(r) d\vec{x}_{(1,d-1)}^2 + dr^2
\ee
But allowing for nontrivial dilaton and Kalb-Ramond background. The dilaton background
 still preserves $IO(1,d-1)$ invariance because it depends on the holographic
coordinate only, $\Phi=\Phi(r)$. 
But any nontrivial antisymmetric background necessarily
breaks it (in the sense that $\pounds(k)H_{ABC}\neq 0$ for all non translational 
Killing vectors $k$); the only remaining unbroken symmetry being the 
 four dimensional translation group, $\mathbb{T}^4$.
\par
Let us begin by assuming an AdS ansatz,
\be
a(r)= e^{2Qr}
\ee
dressing it with a linear dilaton,
\be
\Phi(r)=Qr+\Phi_0
\ee
and allowing for a nontrivial Kalb-Ramond field,
\be
b_{\m\n}(r)=c_{\m\n}e^{2\Phi_0}a(r)+b^{0}_{\m\n}
\ee
where both $c_{\m\n}$ and $b^{0}_{\m\n}$ are constant tensors.
\par
Remarkably enough, this ansatz can easily be shown to be a solution, provided only
that the constants obey the following relationships:
\bea\label{constantes}
&&\eta^{\a\b}c_{\m\a}c_{\n\b}=0\, (\mu\neq\nu\,)\nonumber\\
&&(2-d)\eta_{\m\n}=2e^{4\Phi_0}\eta^{\a\b}c_{\m\a}c_{\n\b}\nonumber\\
&&-d =e^{4\Phi_0}\eta^{\a\delta}\eta^{\b\gamma}c_{\a\b}c_{\delta\gamma}
\eea
These equations are compatible only if
\be
d=\frac{d(d-2)}{2}
\ee
which fixes uniquely
\be
d=4
\ee
which seems worth noticing. On the other hand, the other equations of the set
(\ref{constantes}) are equivalent to
\be
c_{0i}^2 =- c_{jk}^2
\ee
where $i,j,k=1,2,3$ in cyclic order. It is precisely this condition that forbids
a minkowskian solution, if the Kalb-Ramond field is to be real.
\par
Let us define the useful combination 
\be c^2\equiv \sum_i c_{0i}^2 
\ee
The only remaining equation relates this scale with the value of the
dilaton at the origin: 
\be 
c^2 = -4 q e^{-4\Phi_0} 
\ee 
The
gravitational part of this solution is just AdS with radius $R=\frac{1}{Q}=\sqrt{\frac{2\a'}{7}}$; whereas the dilatonic
part is exactly the non-critical linear dilaton; to combine the two in
a non-critical background it has been necessary to turn on the
antisymmetric field, which in turn is only possible in Euclidean
signature. \footnote{A similar $AdS$ solution with B-field was found in \cite{Boer} }
\par
\begin{figure}[h]
\begin{center}
\leavevmode
\epsfxsize=10cm
\epsffile{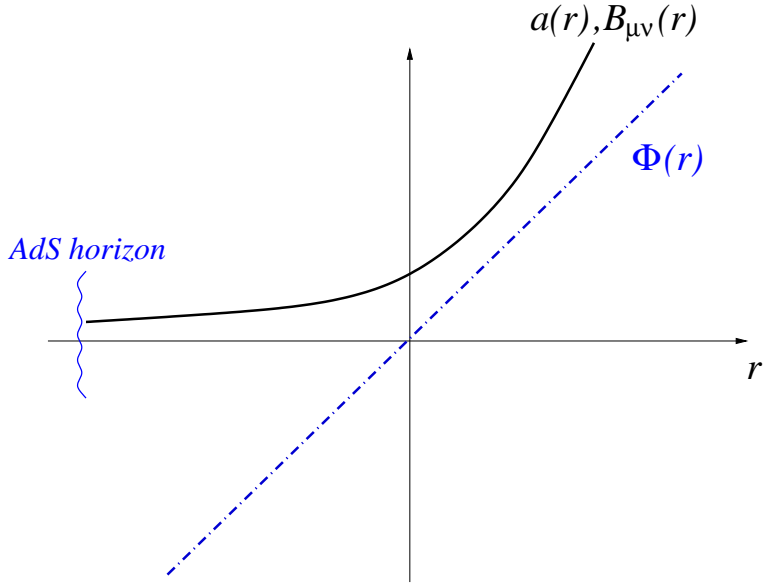}
\caption{\it Warp factor, dilaton and Kalb-Ramond field solution v.s. holographic coordinate}
\label{fig1}
\end{center}
\end{figure}    
There are some simple generalizations of this solution. First of all, we have
considered for the time being the bosonic string only, but nothing prevents us
to assume world sheet supersymmetry (wss). This changes the value of $q$ only, 
to $q=\frac{d-9}{4\a'}=-\frac{5}{4\a'}$, in such a way that the new radius
of AdS is $\sqrt{\frac{4\a'}{5}}$.
\par
A different generalization involves adding a Ricci flat internal manifold
coordinatized by $y$:
\be
ds^2= a(r) d\vec{x}_{d}^2+dr^2+d\vec{y}_{d'}^2
\ee
The new value of the constant $q$ is now $q=\frac{d+d'-25}{6\a'}$, or, with
wss, $q=\frac{d+d'-9}{4\a'}$. It is interesting that by adding
{\em spectator dimensions} the AdS radius becomes larger.
\section{The Interpolating Background}
The solution discussed in the previous paragraph has a strong coupling region,where
the dilaton diverges. It would be interesting to look for a solution which 
somewhat interpolates between the two possible signs of the linear dilaton background
and, as such, has a string coupling constant everywhere bounded.
If we write the general condition for the vanishing of the Weyl anomaly coefficients
(\ref{weyl}) with the previous ansatz we get the equations relating the conformal factor
of the metric with the constant characterizing the Kalb Ramond background:
\bea
&&-\frac{2a a'' +(d-2)(a')^2}{4a} + a'\Phi'+\frac{1}{2a}c^2 e^{4\Phi}=0\nonumber\\
&&-\frac{d}{4}\frac{2aa''-(a')^2}{a^2}+ 2 \Phi''+\frac{dc^2}{4a^2}e^{4\Phi}=0
\nonumber\\
&&q+ (\Phi')^2-\frac{1}{2}\Phi''-\frac{d}{4}\frac{a'}{a}\Phi' + 
\frac{d c^2}{8a^2}e^{4\Phi}=0
\eea
It is a simple matter to check that an euclidean solution is obtained provided $d=4$
and
\be
a(r)=\frac{e^{2Qr}}{1+2Q c' e^{2Qr}}
\ee
where $c'$ is a new constant with dimension of a length. The dilaton is given by:
\be
\Phi=\Phi_0 + log \,a-Qr
\ee

\begin{figure}[h]
\begin{center}
\leavevmode
\epsfxsize=10cm
\epsffile{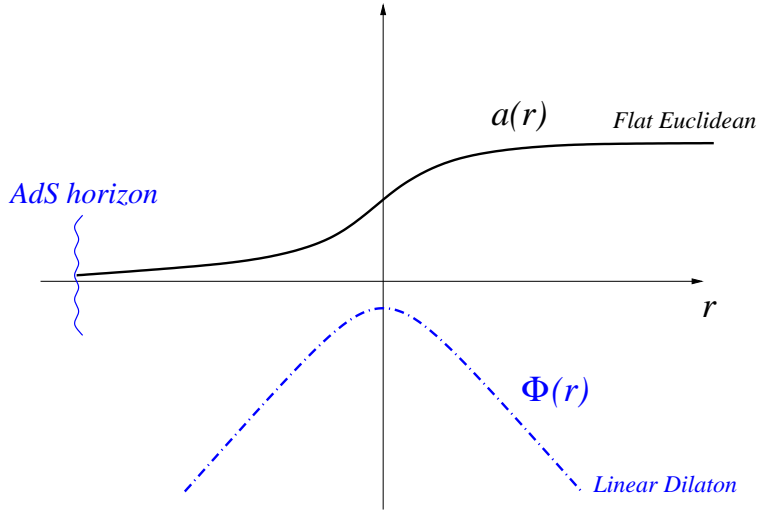}
\caption{\it Warp factor and dilaton of the interpolating background}
\label{fig2}
\end{center}
\end{figure}    

and the Kalb Ramond field strength is just
\be
H_{4\m\n}=2Q\frac{c_{\m\n}}{c} a^2\,e^{-2Qr}
\ee
with $c$ defined above in Equations (16) and (17).
\begin{figure}[h]
\begin{center}
\leavevmode
\epsfxsize=10cm
\epsffile{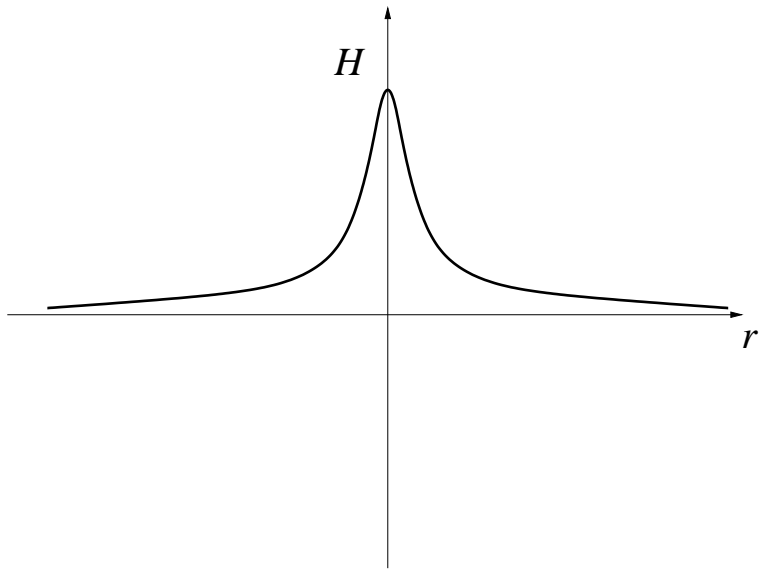}
\caption{\it Kalb-Ramond field strengh shape}
\label{fig3}
\end{center}
\end{figure}

This new solution reduces to the noncritical AdS background of the previous 
paragraph when $c'=0$. On the other hand, when $c'\neq 0$ the dilaton background
is of the interpolating type, because
\be
lim_{r\rightarrow\pm\infty}\Phi(r)=\mp Qr
\ee
Several characteristics of the solution are summarized in the Figures 2,3 and 4.
Notice, in particular, that each component of the Kalb Ramond field strength
has a gaussiam form, with the location of the maximum fixed at $\bar{r}\equiv
-\frac{1}{4Q}log \left( 4Q ^{2}  (c')^2\right)$, and height $\bar{H}_{\m\n}
\equiv \frac{c_{\m\n}}{4cc'}$.
\par

\begin{figure}[h]
\begin{center}
\leavevmode
\epsfxsize=10cm
\epsffile{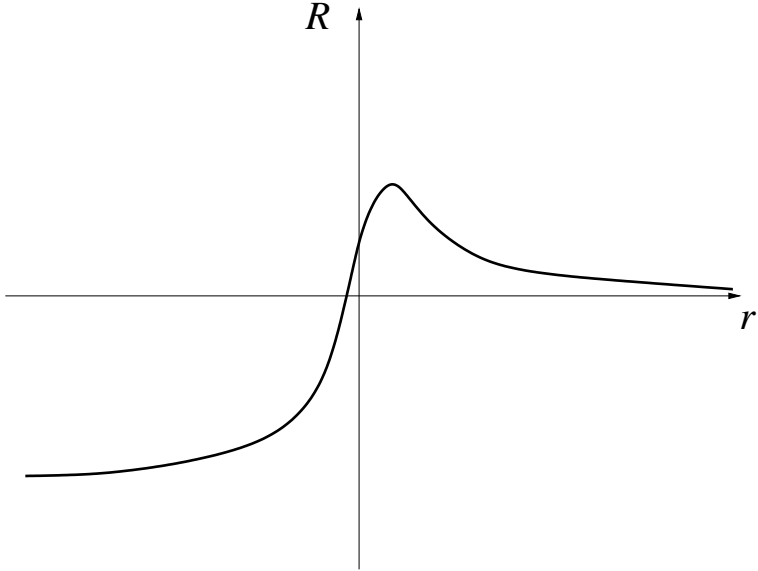}
\caption{\it Curvature scalar. Transition between AdS and flat space is observed}
\label{fig4}
\end{center}
\end{figure}    

There are two features worth discussing. The first is that the solution is 
singularity free. If the scalar curvature is plotted (cf. Figure 4), it can be easily
 seen that there is a maximum curvature $R_{max}=\frac{16Q^2}{9}$ as well as a negative
 minimum value, 
given by $R_{min}= - 20 Q^2$: 
\par

This means that there is some hope of not to be led astray by 
world-sheet sigma model computations. Higher orders in $l_s$ could, however, become 
important as soon as the curvature is big when measured in string units (that is,
$R l_s^2 \geq 1$).
\par
The second comment refers to the form of the graph of the dilaton in Figure 2.
This means that the construct 
\be
g_s\equiv e^\Phi
\ee 
which is usually referred to as the {\em string coupling}  is always small
\be
g_s \leq 1
\ee
In spite of the fact that the string coupling is always small, tadpoles at higher genus can modify our solution.

An interesting characteristic of the solution is the B-field action
\be
I\equiv\frac{1}{12}\int dr d^4 x \sqrt{g}e^{-2\Phi}H^2
\ee
which can be easily computed to be
\be
I=\frac{V_4}{c'}
\ee
where $V_4$ is the four dimensional euclidean volume. The corresponding 
{\em density}
 is then a well defined concept, which  vanishes
for the linear dilaton and diverges in the pure AdS  background.
\par
Actually, vanishing of the  Weyl anomaly coefficients leads to
\be
\frac{1}{12}e^{-2\Phi}H^2=- \nabla_A(e^{-2\Phi}\nabla^A\Phi) + 2q e^{-2\Phi}
\ee
This suggests that $I$ can be used as a way to measure the topological  
 change involved in passing from $\Phi=Q r$ at $r=-\infty$ to $\Phi=- Q r$ at
$r = +\infty$. The generalization to world sheet supersymmetry 
 and/or flat spectator dimensions
is straightforward. 
\par
Finally, if a geometric c-function is defined (cf.\cite{Alvarez:2000jb}) the monotonic behavior depicted in Figure 5 is obtained.

\begin{figure}[h]
\begin{center}
\leavevmode
\epsfxsize=10cm
\epsffile{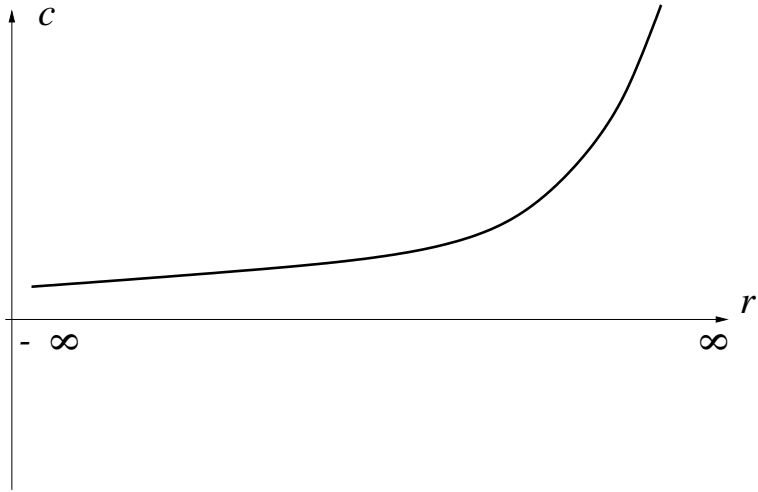}
\caption{\it c-function behaviour} 
\label{fig5}
\end{center}
\end{figure}

\section{Comments on Closed tachyons}

It has often been argued that a closed tachyon condensate is in some sense equivalent
for the string as being non-critical, the precise relationship being given by the
Weyl anomaly coefficient 
\be\label{ta}
q+(\nabla\Phi)^2 -\frac{1}{2}\nabla^2 \Phi -\frac{1}{24}H^2 +\frac{m^2}{16}T^2 =0
\ee
(where $m^2 \a'= - 4$ for the closed string tachyon).
\par
If we assume world sheet supersymmetry (wss), the only changes are in the numerical
value
of q, $q=\frac{D- 10}{4\a'}$  and the tachyon mass, $m^2 \a'= -2$. Then the tachyon 
potential is guaranteed to be even in $T$.
\par
The preceding equation (\ref{ta}) depends only on the combination
\be
\Delta c \equiv q +\frac{m^2}{16}T^2 = \frac{D-6 T^2 -26}{6\a'}
\ee
A vacuum expectation value (vev) for the tachyon is equivalent, from this point 
of view, to a lower dimension. Actually, our solution trivially generalizes
to this case, just by changing $Q\equiv\sqrt{-q}$ by $\sqrt{-\Delta c}$ everywhere.
Of course this does not change the number of geometrical dimensions, which remains
fixed to $d=4$ (plus the holographic coordinate).
\par
In order to anlyze tachyons in $AdS$ we can use the
Breitelohner and Freedman \cite{breitelohner} bound
\be
-\frac{1}{4}(D-1)^2 \leq m^2 R^2
\ee
which for the $AdS_{5}$ solution of section 2  gives 
\be\label{bound}
m^2 \geq -\frac{14}{\a'}
\ee
This is not exactly true, due to dilatonic effects. The relevant lagrangian for a scalar
field in the dilatonic $AdS$ background of section 2 is

\be
S\sim\int d(vol) e^{-2\Phi}\frac{1}{2}(\nabla_A T\nabla^A T - m^2 T^2)
\ee
In our case the $\sqrt{g}$ factor produces an effect $e^{2Qr}$, whereas the dilaton
term
behaves as $e^{-2Qr}$ . 
\par
The explicit form of the action is then
\be
S\sim \int d^4 x dr \frac{1}{2}e^{2 Qr}(e^{-2Qr} (\pd_{\mu}T)^2 + (\pd_r T)^2 -m^2 T^2)
\ee
For  radial fluctuations ($\pd_{\mu} T =0$) the net effect 
is just a change of radius in AdS, from $R = \frac{1}{Q}$, which is the true 
radius of AdS,
to an effective non-dilatonic one, namely $R_e= \frac{2}{Q}$. 
All generic effects of fields propagating in AdS spaces remain the same. Using $R_e$
the bound becomes
\be
m^2 \geq -\frac{21}{6\a'}
\ee
exactly the bound corresponding to the linear dilaton background. As
it is clear from this bound the closed string tachyon of 
$m^{2}=- \frac{4}{\alpha'}$ 
produces an instability.
\par
Let us now assume, for the sake of the discussion, that the tachyon vev complies
with Breitelohner-Freedman bound, that is $T^2\geq 1/2(6)$  
(where the latter figure
stands for the wss case). We then would have a five dimensional $AdS$ background
where the closed string tachyon becomes of positive energy; i.e a {\em 
good tachyon} (in this case the bound (34) becomes $m^2\geq -\frac{4}{\a'}(-\frac{2}{\a'})$ ).
\par
In this framework we can now use 
the interpolating metric to define a tunneling amplitude between both tachyonic vevs. The physical 
effect of the ensuing tunnelling would be to ``restore the symmetry'' in the real vacuum:
\be
<0|T|0> = 0
\ee
This {\em real vacuum}, $|0>$ would be just a quantum mechanical superposition, and, as
such, not a solution of the {\em classical} equations of motion (in the same sense
that the $\theta$-vacuum in QCD is not a classical solution of the Yang-Mills 
equations).
Giving our lack of ability to determine the form of the closed tachyon potential,
the above should be taken as a possible scenario only.

\section*{Acknowledgments}
This work ~~has been partially supported by the
European Union TMR program FMRX-CT96-0012 {\sl Integrability,
  Non-perturbative Effects, and Symmetry in Quantum Field Theory} and
by the Spanish grant AEN96-1655.  The work of E.A.~has also been
supported by the European Union TMR program ERBFMRX-CT96-0090 {\sl 
Beyond the Standard model} 
 and  the Spanish grant  AEN96-1664.The work of L.H. has been supported
by the spanish predoctoral grant AP99-4367460. The work of P.R. has 
been supported in part by a UAM postgraduate grant.


\appendix


\end{document}